# An Approach For Transforming of Relational Databases to OWL Ontology

Mona Dadjoo[1] and Esmaeil Kheirkhah[1]

[1]Department of Computer Engineering, Mashhad Branch, Islamic Azad University, Mashhad, Iran

## ABSTRACT

*Rapid growth of documents, web pages, and other types of text content is a huge challenge for the modern content management systems. One of the problems in the areas of information storage and retrieval is the lacking of semantic data. Ontologies can present knowledge in sharable and repeatedly usable manner and provide an effective way to reduce the data volume overhead by encoding the structure of a particular domain. Metadata in relational databases can be used to extract ontology from database in a special domain. According to solve the problem of sharing and reusing of data, approaches based on transforming relational database to ontology are proposed. In this paper we propose a method for automatic ontology construction based on relational database. Mining and obtaining further components from relational database leads to obtain knowledge with high semantic power and more expressiveness. Triggers are one of the database components which could be transformed to the ontology model and increase the amount of power and expressiveness of knowledge by presenting part of the knowledge dynamically*.

## 1. INTRODUCTION

Today's society is dependent on information systems, so they affect on many of daily routines. Large amounts of information is stored in information systems, they update quickly and make operations act easier. Organizations rely on information systems to manage their operations and interactions with customers and operators, thus they impose a profound impact on society.

Find and integration of information are two main challenges in the information technology areas. Cooperate and interact between different information systems is one of the most important aspects of the daily operations of organizations. Since most of the data and information are in databases especially relational databases, the core of most studies in the field of information is finding and integrating of heterogeneous relational databases. Since relational database systems are designed and implemented separately, numerous challenges are created while these systems should interact with each other.

Applications need to interact with knowledge during data processing. Despite databases are developed a lot, but they are not provided to manage and manipulate data in connection with knowledge. Relational databases cannot use the information that is not listed in the tables explicitly. They don't respond to semantic queries and do not support the inference.

According to support of semantic management in relational databases, other development is needed to eliminate the gap between data and knowledge representation. To solve the question of how to extract semantic information from a database, inference it and obtain valuable information, there is a need to convert the database to the knowledge base.

Along with the development of the semantic web ontology languages like OWL, and applications like web agents and services, it is essential that current data sources such as databases be available for semantic knowledge. So To achieve semantic knowledge it is necessary to make current database content available for these semantic applications which use ontologies to define the meaning of data. Relational databases are currently the largest data sources in the world, but the structure and integrity constraints in relational tables that are defined via schemas are not effective in semantic expressive as ontologies. To resolve the gap between schemas and ontologies and syntactic difference among different describe languages, information systems integration based on ontology is proposed. In this paper, we propose an approach for generating an ontology data model from relational database schema. The main purpose of this paper is transforming special components of database which leads to obtain more enriched ontology. We show that besides the main components of database schema such as tables, fields and relations, further components like constraints and triggers also could be extracted and defined in the ontology model.

The rest of this paper is organized as follows. In Section 2 we present an overview of the methods that have been proposed in connection with the field of transforming relational databases to ontologies. Section3 shows the details of proposed method and specifies the information that must be extracted. Some rules are introduced which database extracted information are transformed to a graph model through them. Further rules and definitions that are necessary to transform graph model to ontology are determined. Finally, Section 4 discusses some conclusions and suggestions for further work.

## 2. Related works

Extract and represent semantic information contained in the database and reasoning and inference of it requires the transformation of relational model to a semantic representation model, ontology model is an appropriate model to response these requirements. Many methods have been provided in connection with transferring relational database to ontology structure that can be classified in different ways. One of the aspects that existing methods can be classified based on it is the type of resource that methods are considered as the source of transmission. In this paper, we divide and offer proposed methods into three general categories: Approaches based on logical model [16],[1],[12],[5],[2],[17], approaches based on conceptual model[47] and approaches based on conceptual middle model[15],[13],[35],[10],[7],[11].

Approaches based on logical model derive their power from database creation language. Since each database maintains its tables' structure in structured query language, a simple way to obtain semantic information could be from this source. [28]

Method which Zhou [17] has provided is placed in this category; in this paper only the information of relational database schema was converted to ontology and data still remain in the database because the author believes that managing and querying data in relational model is easier than the ontology. Despite of converting details of transforming such as constraints which defining characteristics of columns, hierarchy between the extracted concepts were not extracted.

Initiative of the approaches which fall into conceptual model based category is based on this idea that both of the database and the ontology have a conceptual model and transforming of one conceptual model to other is easier and could preserve more semantics. [28]

Approach proposed by Upadhyaya et al. [47] falls in to this category and is introduced with the aim of transfer EER diagram to OWL ontology. The proposed algorithm has focused on capturing conceptual semantics from diagram to build equivalent ontology. The system requires a domain expert to aid more meaningful information and obtain a richer ontology. The main advantage of the method is to cover some cases such as many to many relationships and inheritance. But there are drawbacks such as: unavailability of diagram which in most cases diagram isn't built during designing the database. Despite the availability, the changes which are imposed on the database model have no effect on the diagram. Being dependent on the domain expert is another drawback. One of the most important aspects in converting relational database into ontology is the potential of explicit modeling of information that are modeled implicitly or not displayed in relational databases at all. The difference between logical and conceptual model is that in conceptual model there is a clear distinction between concepts and relations while in logical model tables are used to represent both concepts and some relationships between concepts cases. So just conceptual model expresses inheritance hierarchy obviously. Cardinality constraints are rich in conceptual model; against in relational schema cardinality is displayed implicitly. [28] Conceptual data model is closer to ontology design semantics. On the other hand relational data model that represent by SQL-DDL is the best model for describing properties of entities attributes and data storage. At the same time transforming system needs a reliable source and here reliability is consist of two conditions: one is accessibility of resource and other is the ability of representing the current status of database. These two conditions are just in SQL-DDL schema resource. So logical model which is written in SQL-DDL is an appropriate source to represent database model. But to overcome the weakness of this model, conceptual model can be used to enhance and enrich the SQL-DDL model data source results. Thus, a hybrid approach which is providing conceptual middle model using reverse engineering is an appropriate way for transforming. This conceptual model preserving the features of logical model and each time present current status of database. [28]

Methods based on conceptual middle model apply a middle model in the process of converting database to ontology. First, database is converted into the middle model and then ontology is produced through that model. Indeed this middle model is conceptual data model which is obtained through the reverse engineering on the database. Conceptual data model is independent of the physical implementation of database and extracts domain information from relational schema. To implement the conceptual model, graph theory is used for the graph representation of database structure.

In systems which reverse engineering techniques are used, middle conceptual model is suitable for the cases which the database schema is frequently changing and every time there is need to build a new ontology with any change.

In general, the problem of transforming of relational database to ontology is that a relational database schema is not necessarily compatible with ontology schema; Schema is defined depending on use, indeed information can be stored for different applications under the different schemes. This could be the reason for the use of graph model, meaning that in the graph model we could have better understanding of data structure for ontology-based applications.

The middle model layer, which is actually maps the semantics of source database schema is used to detect changes in the conceptual level (Regardless of physical implementation of relational databases). Another advantage of using conceptual middle model is that it is independent of the physical implementation of database and extracts domain information from relational schema. Indeed, by having middle model the ontology production process will be Independent of relational database management system. In this way, ontology construction from middle model will be a onetime implementation process if relational database management system be any of the MySQL, Oracle or SQLServer management systems. This approach can leads to loose coupling, easy maintenance and Independency. Thus, the reverse engineering method could present better schema and this is the advantage of reverse engineering approach.

Santoso et al. [10] use graph space to represent middle conceptual model and the hierarchical structure of concepts in ontology generation. In this structure which is made of set of edges and nodes, nodes represent classes in the ontology that have been made equivalent to database tables. The edges showing the relationships between classes which are obtained from extracted relations of the database structure. In addition of transforming database structure, this method also transforms records of tables to the instances of ontology classes.

The most challenges related with methods proposed in the field of automatic ontology generation from relational database especially in reverse engineering and graph-based methods is the correctness and accuracy of generated knowledge (ontology). The purpose of this study is to develop a framework for generating ontology from relational database using conceptual model which produces an ontology model in OWL structure while keeping relationships and obtain more semantics from the database and thus present richer ontology than the previous methods.

## 3. Proposed method

Databases store large amount of data, but they are not considered as a knowledge representation language. Against, since the ontology has the ability to get semantics provided in a domain, can be applied for the knowledge representation. But Ontology has not the database abilities in data storage and retrieval. So, Ontologies could be used to express the relationships and semantics within a relational database structure. Whole information of schema and its components including the constraints and advanced topics in SQL like triggers have not intended in previous proposed methods until now.

In our proposed method of transforming relational database to the ontology which is based on three steps, the input of the system is a relational database (Written in SQL Data Definition Language) and its output is an ontology model in the OWL structure that will be produced from the knowledge extracted from relational database components. At first step relational database is received as input, then desired information (Metadata) is extracted and database tables are classified based on it. In the second step the graph components are produced and the middle conceptual model is created using information obtained from previous step. In the third step of system the middle graph is received and ontology model will be created and final ontology is generated based on it. In our approach we use and employ the details of previous provided approaches; all transform cases including hierarchical structures, constraints and limitations of tables and columns are putted together and integrated; these cases are displayed in the middle model. Moreover, we also extract triggers for each table which has trigger definition and by imitation of the event based ontologies [52] we define a corresponding event class in the ontology. Also with reasoning at the ontology level in target ontology model which is obtained during transforming process, obtained events will be fired at the ontology side.

## 3.1. Event representation in the ontology

According to the non-static nature of society, knowledge which reflects the real world needs to be updated regularly. Relational databases have mechanisms to update automatically, so a formal model is necessary to represents events in event based systems for the knowledge representation. This could be gained by converting the relational database trigger definitions at the ontology side. To represent this feature of relational database in the form of knowledge, an event class is defined in the obtained ontology from transition process. Since in SQL, there are three types of triggers (Insert, Update and Delete), so if there are any of these triggers in the source database, correspond sub-class (Insert, Update and Delete) is created for the event class in the ontology model. Now during the conversion of relational database to ontology, reasoning in the obtained ontology is done as follows: when each of events (Insert, Update and Delete) associated with one class occurs, operations specified in the trigger's definition of corresponded table to that class will done, also the name of class which fired the event and a time-stamp which displays the event time are recorded in the event class properties. For example consider a table which has delete trigger. By deletion of each record from table, report of operation is done by insertion of deleted record to an audit table. To convert this case in ontology side, a subclass (delete) of event class will be created. Then for each time deletion of an instance of the corresponded class of table which has delete trigger, deleted instance is add to the corresponded class to audit table. The name of class that instance is deleted from it and a time-stamp is also added to the property values of delete subclass of event class.

## 3.2. Database information extraction

Input of this step of proposed system is relational database written in SQL DDL and its output is a series of data sets to deliver the conceptual middle model production step.

### 3.2.1. Definitions

Let $R$ be the set of tables in relational database so it's defined as $R = \{RE, RR\}$, tables are divided into two entity and relational tables sets.

- *RE* is a finite set of entity tables, data are kept in the entity tables.
- *RR* is a finite set of relational tables. Relational tables are only for displaying the relationship between other tables and do not carry data. Relational table acts as intermediate table in displaying many to many relationships between tables, and thereby many to many relationship is broken to one to many relationships. In these tables primary key is composite of entity tables' foreign keys. For each table available in database at first it is studied which it is placed in *RE* or *RR* sets.
- *PKey(R)* is a function that returns the primary key of table.
- *FKey(R)* is a function that returns the foreign keys of table.
- *REF (FK)* returns referenced table which *FK* foreign key refers to it.
- *CD (RE)* is a function that returns data columns of an entity table.
- *UNQ (RE)* returns columns of *RE* table with unique constraint. A unique value is given for the field with unique constraint via each record.
- *NN (RE)* returns columns with not null constraint of an entity table.
- *Tr (RE)* returns the triggers of *RE* table.
- *Tr_Type(tr)* returns type of *tr* trigger which is "delete", "update" or "insert".
- *Tr_REF(tr)* returns the table which operation in *tr* trigger is applied on it.

By extracting information above, the output information from the first step is completed.

### 3.3. The middle graph model creation

At this step conceptual middle model will be produced in the form of graph model via information extracted from previous step.
#### 3.3.1. Definitions

Let *G* be a directed labeled graph which is obtained in the conceptual middle model creation step. This graph is defined as *G = (N, E)* where: *N= {Na,Nc,Ne}* is a finite set of nodes. *Nc* represent a finite set of class nodes depicted as ellipses in graph. *Na* represents a finite set of attribute nodes depicted as box and *Ne* is a set of event nodes depicted as triangles in the *G* graph. *E* is a finite set of labeled edges; each member of *E* is in the form $E_i<N_1, Label, N_2>$. $N_1, N_2 \in N$, indicates the source and destination node of the edge respectively. Each node and edge has a unique name and can adopt type. Set of type values which attribute nodes can take is {key, unique, not null}. Set of types which event nodes can take is {delete, insert, update}. Each edge may or may not have unique type too.

The Rules for transforming relational database schema components to the graph model are as follows:

• **Table Transferring Rules**

Entity tables in relational database will convert to the class node.

$\forall RE_i \in R => Nc_i.Name \leftarrow RE_i.Name$ (1)

Relational table will be converted to two edges in the graph. First edge will created from correspond node of the entity table which the first foreign key refers to, toward the correspond node to the second referenced entity table of relational table. Second edge will be created vice versa.

$\forall RR_i \in R, fk_1, fk_2 \in FKey(R), ref_1 = REF (fk_1), ref_2 = REF (fk_2) =>$
$E_1< ref_1.name, Has\ a, ref_2.name >,\ E_2< ref_2.name, Has\_A, ref_1.name >$ (2)

• **Column transferring rules**

Columns in the Entity tables will convert to the attribute nodes.
$\forall RE_i \in R, cd_j \in CD\ (RE_i) => Na_j.name \leftarrow cd_j.name, E_j< RE_i.name, Has\_Att, Na >$ (3)

Key column is converted to the attribute node with key type.
$\forall RE_i \in R, pk \in Pkey(RE_i) => Na.name \leftarrow pk, Na.Type \leftarrow Key$ (4)

Constraints of the columns are converted into the type of correspond node. So nodes correspond to the columns with unique or not null value constraints, will get unique or not null type.
$\forall RE_i \in R, cd_j \in CD\ (RE_i), cd_j \in UNQ\ (RE_i) => Na_j.name \leftarrow cd_j, Na_j.Type \leftarrow unique.$ (5)

$\forall RE_i \in R, cd_j \in CD\ (RE_i), cd_j \in NN\ (RE_i) => Na_j.name \leftarrow cd_j, Na_j.Type \leftarrow not\ null.$ (6)

Foreign keys of entity table will convert into the edges in the graph. Edges will created from correspond node of entity table toward the referenced table correspond node. If a foreign key has unique constraint, correspond edge will have unique type.

$$\forall RE_i \in R , fk \in Fkey(RE_i) , ref = REF(fk) \Rightarrow E< RE_i.name, Has\_A, ref.name > \qquad (7)$$

$$\forall RE_i \in R , fk \in Fkey(RE_i) , ref = REF(fk), fk \in UNQ (RE_i) \Rightarrow$$
$$E< RE_i.name, Has\_A, ref.name > , E.Type \leftarrow unique. \qquad (8)$$

Foreign key column indicates relationship between two tables; if this column be primary key at the same, there is a hierarchical relationship between them. Thus a hierarchy will be established between two correspond nodes.

$$\forall RE_i \in R, pk=Pkey(RE_i) , fk \in Fkey(RE_i), pk= fk , ref = REF(fk) \Rightarrow E< RE_i.name, IS\_A , ref.name > \quad (9)$$

• **Trigger Transferring Rules**

Trigger which defined in entity table will convert to the event node in graph G. An edge is established from the correspond node to the owner table of trigger, towards created event node, and one other edge is established from the event node to the correspond node of referenced table (table which operation specified in the defined trigger is applied on it). For delete trigger:

$$\forall RE_i \in R , tr \in Tr (RE_i) , ref = Tr\_REF(tr) , delete=Tr\_Type(tr) \Rightarrow$$
$$Ne.name \leftarrow tr.name, Ne.Type \leftarrow delete, E_1< RE_i.name, Has\_event ,Ne >, E_2< tr.name, To, ref.name > \quad (10)$$

Because it is supposed to built a class for each type of trigger in target ontology so we create a node for each trigger in conceptual graph model.

### 3.4. Ontology model creation

In this step the graph model which was obtained in previous step consider as input and an ontology model is created and store in OWL structure based on this graph model.

### 3.4.1. Definitions

Let *O* be the ontology model which will be obtained in this step. So it defined as *O =(Cls,OP,SP,DP,HR,IFP,FP,I)* where :

- *Cls* is a set of concepts or classes in the ontology.
- *HR* is a hierarchy relationship between two concepts.
- *OP* is an ObjectProperty between two concepts.
- *DP* is a DatatypeProperty between a concept and literal value.
- *FP* is a functional property between a concept and literal value, this property is a sub property of DatatypeProperty.
- *IFP* is an InverseFunctionalProperty between two concepts which is a subProperty of ObjectProperty.
- *I* represents a set of instances in this ontology.

The rules for transforming of graph model components to the ontology are as follows:

• **Nodes transforming rules**

The class node in graph will transform to a concept or class in ontology.
$\forall\ Nc_i\ \epsilon\ N => Cls_i.Name \leftarrow Nc_i.Name.$ (11)

All attribute nodes with no type in the graph will transform to DatatypeProperties in the ontology.

$\forall Na_i, Nc_j\ \epsilon\ N, E< Nc_j, Has\_Att, Na_i >, Na_i.Type = \emptyset =>$
$DP_i.Name \leftarrow Na_i.Name, Domain(DP_i) \leftarrow Nc_j.Name.$ (12)

Key type attribute node will transform to a FunctionalProperty and its MinCardinality and MaxCardinality will be equal to one.

$\forall Na_i, Nc_j\ \epsilon\ N, E< Nc_j, Has\ Att, Na_i >, Na_i.Type = key =>$
$FP_i.Name \leftarrow Na_i.Name, Domain(FP_i) \leftarrow Nc_j.Name, FP_i.Cardinality \leftarrow 1.$ (13)

FunctionalProperty will produced from the unique type attribute node, its MaxCardinality will equal to one.

$\forall Na_i, Nc_j\ \epsilon\ N, E< Nc_j, Has\_Att, Na_i >, Na_i.Type = Unique =>$
$FP_i.Name \leftarrow Na_i.Name, Domain(FP_i) \leftarrow Nc_j.Name, FP_i.MaxCardinality \leftarrow 1$ (14)

If one node has not null type, MinCardinality will be equal to one for the correspond property.

$\forall Na_i, Nc_j\ \epsilon\ N, E< Nc_j, Has\_Att, Na_i >, Na_i.Type = not\ null =>$
$DP_i.Name \leftarrow Na_i.Name, Domain(DP_i) \leftarrow Nc_j.Name, MinCardinality \leftarrow 1.$ (15)

• **Edges Transforming Rules**

Edges in graph will transform to relations between concepts in ontology.

$\forall E<Nc_i, Has\_A, Nc_j >=> OP.Name \leftarrow E.Name, Domain(OP) \leftarrow Nc_i.Name, Range(OP) \leftarrow Nc_j.Name.$ (16)

An *IS_A* labeled edge which indicates hierarchy relationship between two class nodes, will transform to HR relation in ontology. Correspond class to the source node of edge will be a subclass of correspond class with edge target node.

$\forall E< Nc_i, IS\_A, Nc_j > => SubclassOf(Nc_j.name) \leftarrow Nc_i.Name$ (17)

Rule below shows that the edges which have been created from relational tables will be converted to the two ObjectProperties which each of them is inverse of the other one.

$\forall E_i < Nc_1, Has\ a, Nc_2 > \epsilon E, \exists E_j< Nc_3, Has\ a, Nc_4 > \epsilon E, Nc_1= Nc_4, Nc_3= Nc_2 =>$
$OP_1.Name \leftarrow E_i.Name, OP_2.Name \leftarrow E_j.Name, Inverseof(OP_2) \leftarrow OP_1.$ (18)

Edge that has unique type will transform to an InverseFunctionalProperty in ontology. Domain and range creation of this property fallows as ObjectProperty rules.

$\forall E_i < Nc_1, Has\ a, Nc_2 > \epsilon E, E_i.Type= Unique => IFP_i.Name \leftarrow E_i.Name.$ (19)

• **Fire events in the ontology**

If there is a node in the event node set of graph model, an "Event" named class will be built in ontology and two DatatypeProperties that are named "Time" and "Agent" will be created with "Event" class domain. "Agent" property shows the class which the event is fired by it, "Time" property shows the time when the desired event is happened. Each event node in graph will transform to a subclass of "Event" class with the same name of event type. More, an ObjectProperty will be created with domain of correspond class to the node that has event and the range of related subclass of event class (insert delete, update). For example for delete:

$\forall$ $Ne_i$ , $Nc_j$ , $Nc_k$ $\epsilon$ $N$ , $E_1$< $Nc_j$ , $Has\_event$, $Ne_i$ > , $E_2$< $Ne_i$, $To$, $Nc_k$ > , $Ne_i.Type = delete$ => $Cls_1.Name="Delete"$ , $Cls_2.Name \leftarrow Nc_j.Name$ , $Cls_3.Name \leftarrow Nc_k.Name$, $SubclassOf("Event") \leftarrow Cls$, $OP.Name \leftarrow Ne_i.Name$ , $Domain(OP) \leftarrow Cls_2$ , $Range(OP) \leftarrow Cls_3$. (20)

By relating these event classes with event owner concepts via ObjectProperty, these event classes just displayed in the ontology model. But to complete transforming the triggers and fire them in target the ontology, some rules at the ontology level are needed to trigger the events in an appropriate time. So events must be fired besides the showing. For this purpose, when the generated event occurred in the ontology, the specified operation in trigger definition will done and the name of class correspond to the table that has trigger will add to the "Agent" Property value in the subclass of event class (insert, delete, update) and the time of event will add to the "Time" Property value. Thus we could attach following state to (20). We suppose that the state bellow is for the time which $i$ $\epsilon$ $I$ that is one of the instances of $Cls_2$ is deleted.

$Cls_1.TimePropertyValue \leftarrow CurrentDate$, $Cls_1.AgentPropertyValue \leftarrow Cls_2.Name, Cls_3.createInstance(i)$ (21)

The output of this step is an ontology model which is able to store in OWL structure. Also to visual displaying of obtained ontology model, one of the ontology tools management and editing like protégé could be used.

## 4. Conclusion

In this paper we present an approach based on transforming relational databases to the ontology model. Therefore a transition system is introduced which accepts a relational database as input, and an ontology in OWL structure is produced after applying transforming rules. The proposed approach is able to provide a transition system to produce a conceptual model from relational database based on graph theory which leads to product of database graph, and also with transforming of the graph obtained, final ontology is generated. The result of this transition system is conversion of SQL Data Definition Language into the OWL ontology structure. With more components involved in the transition process, this method has succeeded to show richer semantics in the target ontology. The transition to the middle model and ontology model is done automatically. With Using conceptual graph model the final output is obtained independent of the physical implementation of database and database management system.

The proposed approach converts triggers of relational tables that are defined in the form of unconditional and those with the complex and conditional definitions could not be converted. One of the aspects that could be considered further in this discussion is transforming of abnormal database structure to the ontology. Another aspect of this algorithm is conversion of the SQL Queries to SPARQL in the ontology side which consequently more complex triggers, functions and stored procedures could be transformed to the ontology model.

## References

[1]  F.Song, G.Zacharewicz, D.Chen, An ontology-driven framework towards building enterprise semantic information layer. Elsevier 2013.
[2]  H.Zhang, X.Diao, Z.Yuan, J.Chun , Y.Huang , EVis : A system for Extracting and Visualizing Ontologies from Databases with Web Interfaces .IEEE 2013.
[3]  J.BAKKAS, M.BAHAJ, Generating of RDF graph from a relational database using Jena API. International Journal of Engineering and Technology , 2013.
[4]  J.BAKKAS, M.BAHAJ, A.Marzouk . Direct Migration Method of RDB to Ontology while Keeping Semantics. International Journal of Computer Applications (0975 – 8887) Volume 65– No.3, March 2013
[5]  L.Yiqing, L.Lu , L.Chen , Automatic Learning Ontology from Relational Schema. IEEE 2012.
[6]  N. GHERABI, K.ADDAKIRI , Mapping relational database into OWL Structure with data semantic preservation IEEE 2012.
[7]  G.Russo, F.Anastasio, A.Pipitone, A.Gentile, R.Pirrone , VEBO: Validation of E-R diagrams through ontologies and WordNet IEEE 2012.
[8]  H.Santoso, S. Haw, Z.Abdul-Mehdi, Ontology extraction from relational database: Concept hierachy as background knowledge Elsevier 2011.
[9]  W.Ahmed , M.Ahtisham Aslam , J. Shen , J.Yong , A Light Weight Approach for Ontology Generation and Change Synchronization between Ontologies and Source Relational Databases . IEEE 2011.
[10]  X.Zhou, G.Xu, L.Liu , An Approach for Ontology Construction Based on Relational Database science Academy 2011.
[11]  S.Yang, Y.Zheng , X.Yang , Semi-Automatically Building Ontologies from Relational Databases IEEE 2010.
[12]  L.Yuzhao, D.Dongxia. RDB-based Approach to Domain Ontology for Contingency Plan IEEE 2010.
[13]  Z.Telnavora , Relational database as a source of ontology creation IEEE 2010.
[14]  Peng Liu1, Xiaoying Wang, Aihua Bao, Xiaoxuan Wang,"Ontology Automatic Constructing Based on Relational Database", Ninth International Conference on Grid and Cloud Computing  2010.
[15]  S.Zhou, Relational Database Semantic Access Based on Ontology . springer 2010.
[16]  I. Astrova1.Rules for Mapping SQL Relational Databases to OWL Ontologies. Springer 2009.
[17]  S.Sane, A.Shirke,Generating OWL Ontologies from a Relational Databases for the Semantic Web . ACM 2009.
[18]  S.Jia,G.Zhang.Ontology-based knowledge extraction for relational database schema IEEE 2009.
[19]  K. M. Albarrak, E. H. Sibley.Translating Relational & Object-Relational Database Models into OWL Models. IEEE 2009.
[20]  I. Astrova, Rules for Mapping SQL Relational Databases to OWL Ontologies, Metadata and Semantics, Springer 2008
[21]  F.Cerbah.Mining the Content of Relational Databases to Learn Ontologies with Deeper Taxonomies.ACM 2008.
[22]  C.ping, H. Lu, C.Bin.Research and Implementation of Ontology Automatic Construction Based on Relational Database . IEEE 2008.
[23]  J.Trinkunas, O.Vasilecas.Building Ontologies from Relational Databases Using Reverse Engineering Methods. ACM 2007.
[24]  J. Sequeda, S.Tirmizi, O.Corcho, D.Miranker. "Survey of Directly Mapping SQL Databases to the Semantic Web" 2011.
[25]  L.ZHANG , Jing LI , Automatic Generation of Ontology Based on Database Journal of Computational Information Systems 2011.
[26]-  S. Upadhyaya and P.Kumar, "ERONTO: A Tool for Extracting Ontologies from Extended E/R Diagrams", in Proceedings of SAC"05 ACM Symposium on Applied Computing, Santa Fe, New Mexico, USA, March 2005.
[27]  http://www.intelleo.eu.
[28]  N. Alalwan, H.Zedan, F.Siewe, "Generating OWL Ontology for Database Integration "Third International Conference on Advances in Semantic Processing 2009.